\documentclass[twocolumn,pre,a4paper]{revtex4}
\usepackage{graphicx}
\newcommand{\ie}{{\em i.e. }}
\newcommand{\eg}{{\em eg. }}
\newcommand{\cf}{{\em cf. }}
\begin{document}

LPT Orsay 04/03
\vspace{0.6cm}

\title{\bf PERTURBING GENERAL UNCORRELATED NETWORKS}
\author{Z. Burda $^1$, J. Jurkiewicz $^1$ and A. Krzywicki $^2$}
\affiliation{$^1$ M.~Smoluchowski Institute of
Physics, Jagellonian University,
Reymonta 4, 30-059 Krakow, Poland\\
$^2$ Laboratoire de Physique Th\'eorique, B\^atiment 210,
Universit\'e Paris-Sud, 91405 Orsay, France\\}

\begin{abstract}
This paper is a direct continuation of an earlier work, where
we studied Erd\"os-R\'enyi random graphs perturbed by an interaction 
Hamiltonian favouring the formation of short cycles. Here, we generalize
these results. We keep the same interaction Hamiltonian but let it act
on general graphs with uncorrelated nodes and an arbitrary given degree 
distribution. It is shown that the results obtained for Erd\"os-R\'enyi
graphs are generic, at the qualitative level. However, scale-free graphs
are an exception to this general rule and exhibit a singular behaviour, 
studied thoroughly in this paper, both analytically and numerically.
\vspace{0.5cm}
\end{abstract}
\maketitle

\section{INTRODUCTION}
The best understood random graphs are those with 
uncorrelated nodes and a local tree structure. It 
was natural to develop the statistical mechanics of 
random networks starting with this highly idealized 
picture, as it is natural to begin explaining the 
properties of gases starting with the ideal ones.
However, short loops show up in most real networks 
relatively frequently and it is evident that current 
models should be upgraded to capture this common trait 
(the present state of art in network research is 
excellently reviewed in refs. \cite{ab,dm,new}). A
possible strategy consists in adding to the  
Hamiltonian a term favouring the appearance of 
specific motifs. This strategy was adopted in our
recent publication \cite{bjk}, where Erd\"os-R\'enyi 
graphs were perturbed by adding to the action a term
proportional to the number of triangles \cite{cb}. The 
purpose of the present paper is to outline a possible 
generalization of the discussion of ref. \cite{bjk} to 
the case where in zeroth order the graphs belong to 
the statistical ensemble of simple (\ie non-degenerate) 
graphs with uncorrelated nodes and an {\em arbitrary} 
degree distribution $p_k$.
\par
The partition function of the perturbed model is written
in the form
\begin{eqnarray}
Z = \sum_M \delta\bigl(Tr(M^2) - 2L\bigr)\; e^{S(M)}
 \prod_{j=1}^N \bigl(p_{k_j} k_j!\bigr) 
\nonumber \\
 = Z_0 \langle e^{S(M)}\rangle
\label{pf}
\end{eqnarray}
where $N$ and $L$ are the number of nodes and links, 
respectively, the sum is over adjacency matrices M,
$k_j$ is the degree of the $j$-th node satisfying
$Tr(M^2)=\sum_j k_j$ and $S(M)$ is the perturbing 
Hamiltonian. Setting $S(M)=0$ one obtains a standard 
model of networks with uncorrelated nodes and the degree 
distribution $p_k$ (provided the relation 
$\sum_k kp_k/\sum_k p_k = 2L/N$ holds, 
see for example refs \cite{bl,bk}). 
In the second line we denote by $\langle \cdots \rangle$ 
the average taken in the unperturbed ensemble and therefore $Z_0$ 
is, obviously, the partition function 
of the unperturbed model, in our 
context an irrelevant overall factor (it can be calculated 
analytically in the large $N$ limit, see \eg \cite{bl}, but 
we do not need this result here).
\par
One can, of course, go over from (\ref{pf}) to a grand-canonical 
ensemble, with fluctuating $L$, multiplying (\ref{pf}) by an 
appropriate $L$-dependent weight and summing over $L$. In the 
Erd\"os-R\'enyi theory there is a very natural and simple 
recipe for the weight. Hence, in \cite{bjk} we were grand-canonical.
Here, we find it more elegant to keep the number of links fixed.
Notice, that we are interested in the large $N$ limit where, for
any reasonable choice of the weight, $L$ stays anyhow close to
its average value as a consequence of the constraint 
$L=\frac{1}{2}\sum_j k_j$ and of the central limit theorem.
\par
As in ref. \cite{bjk} we set for definiteness 
$S(M) = \frac{G}{3!} Tr(M^3)$. Expanding the exponential in
(\ref{pf}) we obtain a perturbative series representing the partition
function. This perturbative representation was thoroughly studied in 
ref. \cite{bjk} and the analytic arguments were completed by 
numerical simulations. Let us 
recall the salient conclusions of this study:
\par
We have introduced a diagrammatic representation of possible
contributions to the perturbative expansion, reminiscent of
Feynman diagrams used in field theory. Each diagram is
a specific subgraph of the full random graph. We have shown 
that the number of contributing diagrams grows factorially 
with the order of the perturbation theory. 
\par
As is well known, such a factorial growth of the number of diagrams 
signals a breakdown of the perturbation theory. Indeed, our numerical
simulations indicate the presence of a transition from a smooth,
perturbative regime to a crumpled phase, where almost all nodes
form a complete clique \cite{str}. The transition point $G_{out}(N)$ 
scales like $\ln{N}$. The two phases are separated by a "barrier"
which becomes impenetrable when $N \to \infty$.
\par
Remarkably enough it is possible to 
sum up the leading diagrams, \ie those 
whose contribution survives in the limit 
$N \to \infty$. Thus, for example, 
we have been able to derive a closed analytic expression for the
average number of triangles. It turns out, that at large enough
$N$ this analytic formula is a very good approximation in the almost
whole region $G < G_{out}(N)$. Setting $G=G_0 \ln{N} < G_{out}(N)$ 
one obtains a network model with a nontrivially behaving clustering 
coefficient $C \propto N^{G_0-1}$. This clustering coefficient is 
never constant, it falls to zero as $N \to \infty$, but this fall 
can be made fairly slow by a proper choice of $G_0$.
\par
The results summarized above are the starting point of the present paper, 
which is a direct continuation of ref. \cite{bjk}. In Sect. II we 
extend our diagrammatic rules to general uncorrelated graphs with 
a given degree distribution. We also show that the results of ref. 
\cite{bjk} continue to hold, only slightly modified,  in this 
generalized set-up. The so-called scale-free graphs are the only 
exception to the generic behaviour and require a separate discussion, 
presented in Sect. III. Analytic results are confronted to numerical 
simulations in Sect. IV. We conclude in Sect. V.

\section{EXTENDING THE DIAGRAMMATIC RULES}
The perturbation series is defined as in \cite{bjk}. One calculates the 
successive terms in the expansion
\begin{equation} 
\langle e^{\frac{G}{3!} Tr(M^3)}\rangle = \sum_n \frac{G^n}{6^nn!} 
\langle[ Tr(M^3)]^n \rangle
\end{equation}
This boils down to the calculation of the expectation value of strings
like $M_{a_1a_2}M_{a_2a_3} \cdots M_{a_{3n-1}a_{3n}}$. 
A string does not vanish iff all the matrix elements it involves 
are equal to unity. The only problem 
is that the same element, or its transpose, 
can appear several times in the 
same string and one needs, therefore, 
to catalogue all possible string 
structures (\cf ref. \cite{bg}). This 
is done with the help of diagrams:
\par 
Each matrix element $M_{ab}$ is represented 
by a line segment with endpoints
$a$ and $b$. A string is then represented 
by a collection of $n$ triangles,
possibly glued together. In order to 
calculate the perturbation series one
has to consider all possible diagrams. 
The meaning and the construction of
diagrams is explained at length in ref. 
\cite{bjk}, with the help of explicit
examples. We will not repeat this 
discussion here, referring the reader to the
original paper. We wish only to insist on the salient steps:
\par
(a) All $n$-th order terms of the series are, 
of course, proportional to the
common factor $G^n/6^nn!$.
\par
(b) Every diagram is a subgraph which has 
to be embedded in the full graph.
An $n$-th order diagram has, say, $v$ 
vertices and $v \leq 3n$. These $v$
vertices can be identified with graph nodes in 
$N!/(N-v)! \sim N^v$ manners.
\par
(c) The same diagram topology usually 
represents a number of distinct strings. 
The calculation of this number is the 
relatively difficult part of the game.
But it is universal, in the sense that 
it does not depend on the degree 
distribution. The calculations of 
ref. \cite{bjk} were done for the 
Erd\"os-R\'enyi graph, but hold quite generally.
\par
(d) Finally, one has to find the expectation 
value of the string corresponding
to a given diagram. In Erd\"os-R\'enyi 
theory this is simple: if a diagram
has $\ell$ edges, then there are $\ell$ 
independent adjacency matrix elements
in the string and the expectation value 
is just $p^{\ell}$, where $p$ is
the control parameter equal to $p=\alpha/N$ 
when the average degree is set to 
be finite. The case of a general 
model of uncorrelated graphs with a given 
degree distribution requires some extra 
thought. We will use an argument which 
is not quite original, since it has 
already been employed by other authors 
in a somewhat different context (see, 
for example, refs. \cite{ffh,bo,itz})~:
\par
The probability that nodes $a, b$ are connected is
\begin{equation}
Prob(a,b)=\frac{k_a k_b}{\langle k \rangle N}
\label{ansatz}
\end{equation}
where $k_a (k_b)$ is the degree of a-th (b-th)node and 
$\langle k \rangle = \sum_k kp_k/\sum_k p_k$. Notice that 
$\langle k \rangle N$ equals the number 
of directed links. The probability 
in question is inversely proportional 
to the total number of directed links 
and is proportional to the degrees 
of the nodes. Eq. (\ref{ansatz}) holds
when the right-hand side is small enough, \ie at large $N$.
The probability that $b$ is in turn 
connected to, say, $c$ is however
\begin{equation}
Prob(b,c\mid a)=\frac{(k_b -1) k_c}{\langle k \rangle N}
\end{equation}
because one link emerging from $b$, the one connecting $b$ to $a$, is
already occupied: only $k_b -1$ links 
are potentially "active". Pursuing
the argument one derives the following rules~:
\par
(a) a factor $1/(\langle k \rangle N)$ is associated with every edge of
the diagram, and
\par
(b) a factor $k_a!/(k_a-m_a)!$ is associated with every vertex, say $a$, 
of the diagram: here $k_a$ is the degree of the $a$-th graph node and 
$m_a$ is the degree of the same node regarded as the diagram vertex.
\par
This was for a particular graph. Averaging, one is led to replace 
\begin{equation}
k_a!/(k_a-m_a)!  \rightarrow \langle k_a!/(k_a-m_a)! \rangle
\label{ansatz2}
\end{equation}
in the above rules given above \cite{foot}.
\par
We have assumed here that node degrees 
are uncorrelated, which strictly 
speaking is only true in the limit 
$N \to \infty$. Even in a so-called 
uncorrelated network some "kinematic" 
correlations appear when one imposes 
the constraint that there are no self 
and multiple connections between 
nodes and when $k^2/N$ is not always negligible \cite{bk}. The last 
condition is easily satisfied when the degree distribution is defined 
on a finite support, but may be jeopardized in scale-free networks.
\par
Let us check that for Erd\"os-R\'enyi graphs one gets the result of
ref. \cite{bjk}: when the degree distribution is Poissonian the
average of a binomial moment is just a power of $\alpha$~:
\begin{equation}
\langle k_a!/(k_a-m_a)! \rangle = \alpha^{m_a}
\end{equation}
Of course, one has
\begin{equation}
\sum_a m_a = 2\ell
\end{equation}
Furthermore, in Erd\"os-R\'enyi theory 
$\langle k \rangle = \alpha$. Hence 
the diagram with $\ell$ vertices gets a factor
\begin{equation}
\alpha^{2\ell}/(\alpha N)^\ell = p^\ell
\end{equation}
This is exactly what one has in the grand-canonical Erd\"os-R\'enyi
ensemble and also what one expects in our setup in the large $N$ limit.
\par
We are now equipped to calculate the 
contribution of an arbitrary diagram.
For example, in the limit $G \to 0$  
the average number of triangles in 
a graph, the derivative of the free 
energy with respect to $G$, equals
\begin{equation}
\langle T \rangle_{G=0} = \frac{1}{6}
\left(\frac{\langle k(k-1)\rangle}{\langle k\rangle}\right)^3 
\label{T0}
\end{equation}
because the diagram has three vertices 
of order two. Hence, each vertex
contributes a factor $\langle k(k-1)\rangle$ - 
remember that vertices are
independent - while each link contributes
a factor $1/\langle k\rangle$. The powers of 
$N$ cancel, as in the calculation
of ref. \cite{bjk} and the only difference 
is that now instead of $\alpha$ 
appears a ratio of binomial moments of the degree distribution $p_k$. 
\par
Higher order diagrams call for other binomial 
moments. It is important to 
realize that as long as all the moments of $p_k$ are finite - and 
therefore $N$-independent for large enough network size - 
the hierarchy of diagrams in the 
$\frac{1}{N}$ expansion is the same as in
ref. \cite{bjk}. In particular, the same 
diagrams are leading. Summing the
leading diagrams, \ie those whose 
contribution remains finite when $N \to \infty$,
one gets
\begin{equation}
\langle T \rangle = \frac{1}{6} 
\left(\frac{\langle k(k-1)\rangle}{\langle k\rangle}\right)^3 e^G
\label{tvsG}
\end{equation}
which generalizes eq. (36) of ref. \cite{bjk}. 
Higher order moments (cumulants) of the 
$T$-distribution are obtained by differentiating the 
right-hand side of (\ref{tvsG}) with respect 
to $G$. One finds that the distribution is Poissonian.

\par
The situation changes when the moments 
of $p_k$ are not necessarily finite.
Without much loss of generality we will 
limit our attention to the so-called
scale-free graphs, \ie those where 
$p_k$ falls at large $k$ like a power:
$p_k \propto 1/k^{\beta}$. We also assume 
that $\langle k\rangle$ is finite: 
$\beta > 2$.

\section{SCALE-FREE GRAPHS}
When $p_k \sim a/k^{\beta}$ at $k \gg 1$, the moments of order larger 
than $\beta - 2$ diverge. 
Actually, at large but finite $N$, the degree distribution of 
a non-degenerate uncorrelated
graph is cut at 
\begin{equation}
k_{max} \propto N^\gamma, \; \; \;
\gamma = \min \bigl(1/2, 1/(\beta -1)\bigr)
\label{cut}
\end{equation}
(see ref. \cite{bk} for a derivation of 
this result). Hence, higher order moments
of the degree distribution increase like 
some powers of $N$ and, consequently, 
the hierarchy of the $\frac{1}{N}$ expansion 
is modified compared to ref. 
\cite{bjk}. It turns out that three cases 
require a separate discussion:

\begin{figure}
\vspace{-1cm}
\includegraphics[width=6cm]{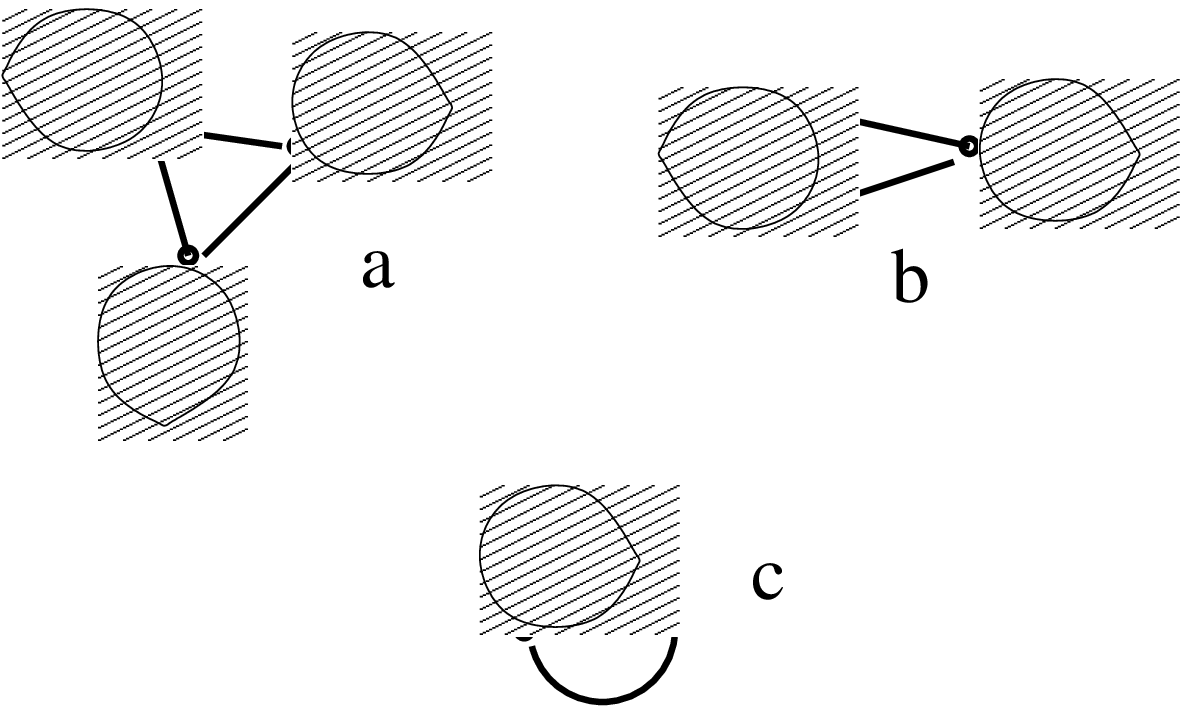}
\caption[Fig.1]{\label{fig1}
Additions of a triangle to a diagram, leading to the increase of 
the number of diagram links by three (a), two (b) and one (c).}
\end{figure}

\subsection{Case $\beta > 3$}
We will prove that the leading diagrams are the same as those 
considered in ref. \cite{bjk}, \ie those where the number of 
triangle edges is equal to the number of triangle vertices. 
One can easily see that in these diagrams the triangles can 
overlap, but otherwise do not touch.
\par
The proof is by induction. First we 
observe that the diagram of order $n=1$ (one triangle) is leading and
its contribution tends to a constant as $N \to \infty$. Higher
order diagrams can be constructed by adding successive triangles.
Suppose that at the $n$-th order the leading diagrams are those with
the number of links equal to the number of triangles. What happens
when one adds the $(n+1)$-st triangle ? If it is put on top of an
existing triangle or if it is isolated, the new diagram scales with 
$N$ the same way as its ancestor. In general, the new triangle is not 
isolated and its addition 
increases the number of links. Three possibilities,
illustrated in Fig. 1, can arise. Because of (\ref{cut})
increasing the degree of a diagram vertex by $\Delta m$ links
produces at most an extra divergent factor 
\begin{equation}
N^{\frac{\Delta m}{\beta-1}}
\end{equation}
When the number of links is increased 
by three (Fig. 1a) $\Delta m =2$
in three vertices. The new links 
yield a factor $N^{-1}$ each. Thus
the new global factor is at most
\begin{equation}
N^{\frac{6}{\beta-1} - 3} = N^{\frac{9-3\beta}{\beta-1}}
\end{equation}
Similarly, for Figs. 1b,c
one finds that the potentially dangerous new factor is at most
\begin{equation}
N^{\frac{6-2\beta}{\beta-1}} \; \; \mbox{\rm and } \; \;
N^{\frac{3-\beta}{\beta-1}}
\end{equation}
\noindent
respectively. For $\beta > 3$ these factors do not diverge. 
We conclude that the leading diagrams are those
considered in ref. \cite{bjk}. The average number of
triangles is finite in the limit $N \to \infty$ and is
given by eq. (\ref{tvsG}). 

\subsection{Case $\beta = 3$}
As already recalled, at finite $N$ 
the degree distribution is cut and
the cut-off scales like $\sqrt{N}$. Of 
course, the cut-off is not sharp, 
but in calculations it is convenient 
to replace it by a sharp one 
$k_{max} = c\sqrt{N}$. The determination 
of the constant $c$ is not 
obvious. We make a choice which yields 
more or less correct moments of 
the cut degree distribution: at large enough $N$ 
\begin{equation}
\langle k^2 \rangle \approx \frac{a}{2} \ln{N}
\end{equation}
and
\begin{equation}
\langle k^m \rangle \approx \frac{ac^{m-2}}{m-2}N^{\frac{1}{2}(m-2)} 
\; , \; \; m > 2 
\label{mom}
\end{equation}
(the sharp cut-off upsets the normalization, 
but the effect is negligible 
for large $N$). The constant $c$ 
can be roughly estimated using 
eq. (\ref{mom}):
\begin{equation}
c \approx \frac{\bigl(\frac{m-2}{a}
 \langle k^m \rangle\bigr)^{\frac{1}{m-2}}}{\sqrt{N}}
\label{c}
\end{equation}
Let us calculate the leading 
$N$-dependence of an arbitrary diagram 
with $\ell$ edges and $v$ vertices: one 
has a global factor (we omit logs)

\begin{equation}
N^{v-\ell + \frac{1}{2}\sum_a (m_a-2)} ,
\label{glo}
\end{equation}

\begin{figure}
\includegraphics[width=5cm]{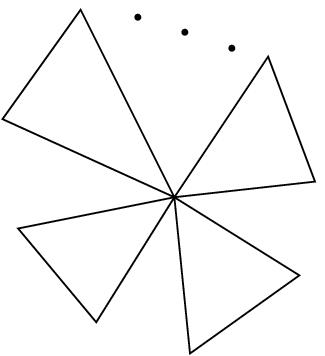}
\caption[Fig.2]{\label{fig2}
Leading diagram for $\beta=3$.}
\end{figure}

\noindent
since each link yields $N^{-1}$, 
the $v$ vertices yield $N^v$ and 
the binomial moments associated with 
$v$ vertices yield the sum in the exponent. 
However, this sum can be calculated: 
$\sum_j (m_j-2) = 2 \ell - 2 v$. Hence, 
the total exponent is zero, only logs
remain. 
\par
As explained in ref. \cite{bjk}, calculating 
the average number of triangles 
at large $N$ it is sufficient to 
consider connected diagrams only. At each order of 
perturbation theory the leading connected diagram is 
the one with the largest number of vertices 
of order 2, \ie the diagram shown 
in Fig. 2. In the $n$-th order we 
have $n$ triangles and therefore the number
of vertices of order $2$ is $2n$. The 
central vertex has order $2n$. 
Each triangle in Fig. 2 can be glued 
to the central vertex in three manners. 
The contribution of the $n$-th order 
diagram to the free energy is therefore (for $n>1$):
\begin{eqnarray}
\frac{G^n}{6^nn!}\; 3^n\; \frac{ac^{2n-2}}{2n-2}\; 
\frac{1}{\langle k \rangle^{3n}}\; 
\langle k(k-1)\rangle^{2n}  \nonumber \\    
\sim \frac{G^n}{2(n-1)n!}\;
 \frac{a^{2n+1}c^{2n-2}}{(8\langle k \rangle^3)^n}\;
\ln^{2n}{N}
\end{eqnarray}
\noindent
where we have used the the asymptotic result
\begin{equation}
\langle k(k-1)\rangle \sim \frac{a}{2} \ln{N}
\end{equation}
In this limit eq. (\ref{T0}) reads

\begin{equation}
\langle T \rangle_{G=0} \sim 
\frac{a^3}{48\langle k \rangle^3} \ln^3N
\end{equation}

Summing all the connected diagrams one gets the free energy. 
Differentiating the latter with 
respect to $G$ one finds the average
number of triangles, which can be written in the form
\begin{equation} 
\langle T \rangle = T_0 + 
\frac{a(6T_0)^{\frac{2}{3}}}{4\langle k \rangle}
\sum_{n=1}^\infty \frac{\bigl(\, b\, 
 G\, T_0^{\frac{2}{3}}\bigr)^n}{nn!}_{\;
\; \stackrel{\mbox{\large $\sim$}}{T_0 \to \infty}} \; 
\frac{a}{2 c^2 G}e^{b G T_0^{\frac{2}{3}}}
\label{ttt}
\end{equation}
where $T_0 \equiv \langle T \rangle_{G=0}$ and
\begin{equation}
b = \frac{c^2\; 6^{\frac{2}{3}}}{2\langle k \rangle}
\end{equation}
It is evident that the increase of $\langle T \rangle$ with 
$G T_0^{\frac{2}{3}} \propto G \ln^2N$ 
is nearly exponential.

\subsection{Case $2 < \beta < 3$}
For $\beta < 3$ all the moments, apart 
from the first, diverge like a power 
of $N$. As shown in \cite{bk} the 
cut-off of the degree distribution scales 
again like $\sqrt{N}$. Therefore an 
arbitrary diagram's contribution is 
proportional to the factor
\begin {equation}
N^{v - \ell + \frac{1}{2}\sum_j (m_j-\beta+1)}
 = N^{\frac{v}{2}(3-\beta)}
\end{equation}
Hence the leading diagrams are those 
with a maximal number of vertices. One 
can calculate the number of triangles 
$\langle T\rangle$ in the leading-power 
approximation. Again, it is sufficient to 
consider connected diagrams only. 
Assuming here, for simplicity of writing, that 
$\langle k^m\rangle = c^m N^{\frac{1}{2}(m+1-\beta)}$ and using 
the combinatorial coefficients calculated in ref. \cite{bjk} we get:
\begin{eqnarray}
\langle T \rangle = \frac{1}{6}\; \frac{c^3}{\langle k\rangle^3}\; 
N^{\frac{3}{2}(3-\beta)} +
\frac{1}{4}\; \frac{Gc^5}{\langle k\rangle^6}\; 
N^{\frac{5}{2}(3-\beta)}  \nonumber \\
+ \frac{7}{16}\; \frac{G^2c^7}{\langle k\rangle^9}\; 
N^{\frac{7}{2}(3-\beta)} + \dots
\label{uuu}
\end{eqnarray}
It is obvious that the result is very 
sensitive to the value of $c$. This
would also be true if we were using 
a more realistic ansatz, viz. the 
analogue of (\ref{mom}). A much more 
reliable prediction is the approximate 
scaling law:
\begin{equation}
\langle T \rangle = T_0 f(G T_0^{\frac{2}{3}})
\label{scaling}
\end{equation}
In the large $N$ limit eq. (\ref{T0}) reads
\begin{equation}
T_0 \equiv \langle T \rangle_{G=0} \propto N^{\frac{3}{2}(3-\beta)}
\end{equation}
and $f(x): f(0)=1$ is a positive, monotonically increasing function. 
Although we have calculated only the first few terms in the expansion
(\ref{uuu}), an educated guess is that the rise of the right-hand side
is again nearly exponential.
 
\begin{center}
\begin{table}
\caption{The average number of triangles 
at $G=0: \langle T\rangle_{G=0}$
from Monte Carlo simulation (left) and 
estimated from degree distributions
using eq. (\ref{T0}) (right). \label{tab1}}
\vspace{0.4cm}
\begin{tabular}{|r|lr|lr|lr|} \hline
\multicolumn{1}{|c|}{$N$} & \multicolumn{2}{|c|}{$\beta=2.5$} 
&\multicolumn{2}{|c|}{$\beta=3$} & 
\multicolumn{2}{|c|}{$\beta=4$}\\ \hline
    &  &  &  &  &  &   \\ 
$1024\;$ &  $\; 26.36(9)$ & $35.49 \;$ &  $\; 11.01(3)$ & $13.15\;$ & 
$\;4.91(2)$ &  $5.31\; $ \\
&  &  &  &  &  &  \\ \hline
    &  &  &  &  &  &   \\
$2048\;$ &  $\; 42.26(18)$ & $54.91\;$ &  $\;15.10(7)$ & $17.58\;$ & 
$\;5.80(2)$ & $6.14\;$ \\ 
&  &  &  &  &  &   \\ \hline
    &  &  &  &  &  &   \\
$4096\;$  & $\; 67.05(36)$ & $84.16\;$ & $\; 20.26(12)$ & $22.93\;$ &
$\; 6.58(3)$ & $6.85\;$ \\
&  &  &  &  &  &   \\ \hline
    &  &  &  &  &  &   \\
$8192\;$  & $\; 106.2(9)$ & $129.5\;$ &  $\; 27.22(28)$ & $30.28\;$ & 
$\;7.33(3)$ &  $7.56\;$ \\
&  &  &  &  &  &   \\ \hline
    &  &  &  &  &  &   \\
$16384\;$ & $\; 167.0(1.2)$& $198.9\;$ & $\; 35.54(33)$ & $38.80\;$ &
$ \; 8.01(9)$ & $8.19\;$ \\
&  &  &  &  &  &   \\ \hline
    &  &  &  &  &  &   \\
\multicolumn{1}{|c|}{$\; N \to \infty\;$} &
\multicolumn{2}{|c|}{$\frac{c^3}{48}\; N^{\frac{3}{4}}$} &
\multicolumn{2}{|c|}{$\frac{1}{6}\; \ln^3N$} &
\multicolumn{2}{|c|}{$\frac{\mbox{\small 32}}{\mbox{\small 3}}$}\\ 
&  &  &  &  &  &   \\ \hline
\end{tabular}
\end{table}
\end{center}

\section{NUMERICAL RESULTS}
In this section we continue the study 
of the preceding one but having recourse 
to Monte Carlo simulations. As in ref. 
\cite{bjk} we use the algorithm of ref. 
\cite{bk}, which has the advantage of 
generating not only nondegenerate graphs 
but also thermal fluctuations. The latter 
point is important because our main 
goal is to check the stability of the 
smooth, perturbative phase. We will also 
compare the numerical Monte Carlo data 
to the predictions of the preceding 
section, obtained by summing leading 
diagrams. The agreement will be only 
semi-quantitative, because of large 
finite-size corrections.
\par
Actually, three types of effects 
occur at finite $N$: First, eqs. 
(\ref{ansatz})-(\ref{ansatz2}) 
are only approximate, especially 
when the degree
distribution has a fat tail, as 
already explained. Second, nonleading
diagrams are negligible, at fixed $G$, 
only asymptotically and it turns out 
that asymptopia is hard to reach. 
Furthermore, the nonleading contribution 
blows up as one approaches the transition 
point $G=G_{out}$, because the 
transition is an intrinsically non-perturbative 
phenomenon. These two effects 
are in a sense a nuisance for us, since they 
just obscure the picture. Third, 
there are manifestations of the fat tail 
$p_k \sim k^{-\beta}$ in the node degree distribution
that persist at any $N$ and are, 
in fact, increasingly important as $N$ 
grows up. They are responsible for 
the singular behaviour of scale-free 
networks and are therefore of much 
physical interest.

\begin{figure}[bbb]
\includegraphics[width=6cm]{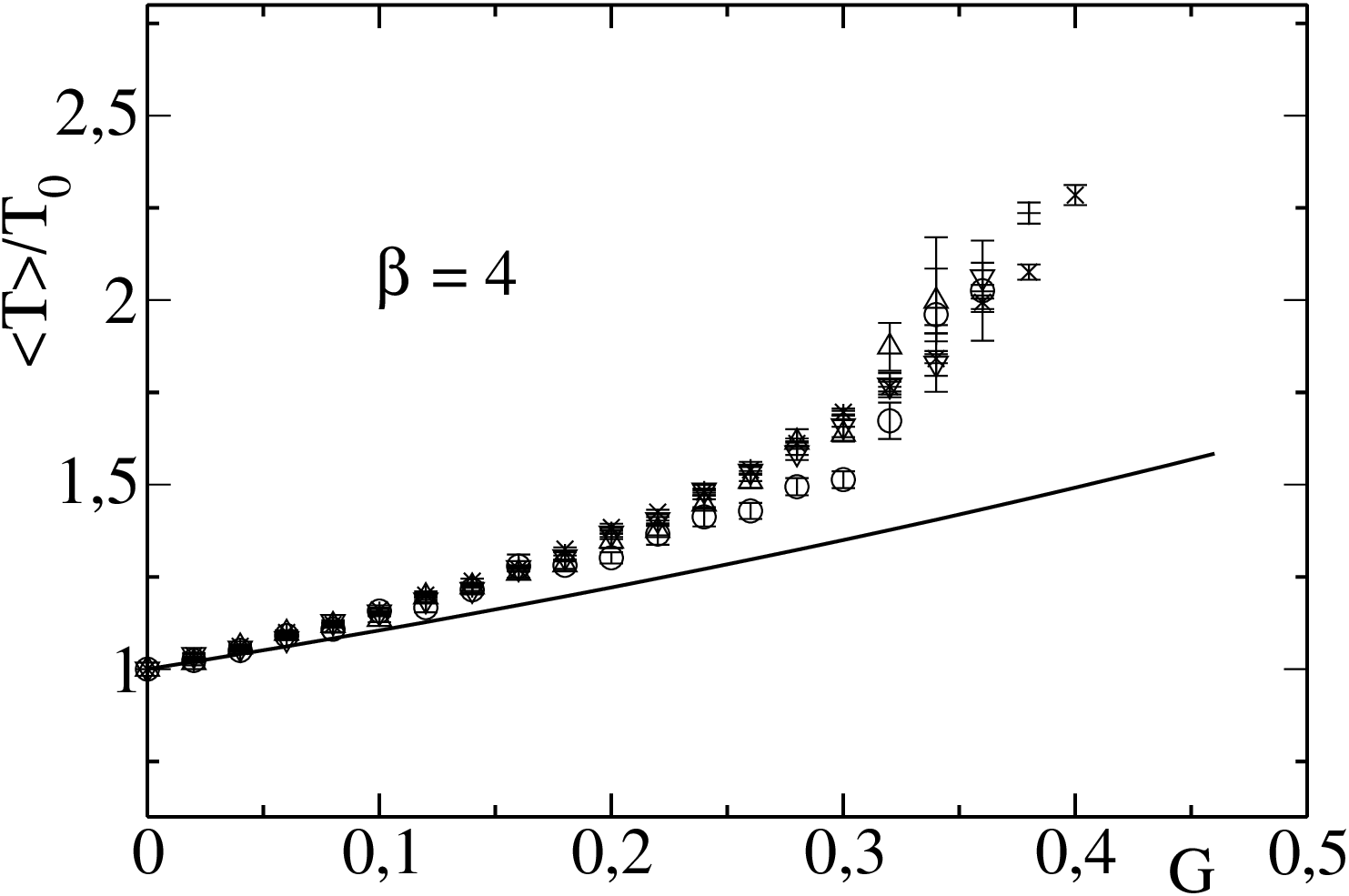}
\caption[Fig.3]{$\langle T\rangle/T_0$ versus
$G$ for $\beta=4$ and $N=2^{10}\; (\times)$,
$2^{11}\; (+)$, $2^{12}\;
 (\bigtriangledown)$, $2^{13}\; (\bigtriangleup)$,
$2^{14}\; (\bigcirc)$.}
\end{figure}
\par
As in our earlier publications we assume, 
for definiteness, that the degree 
distribution has the form (\cf ref. \cite{kr})
\begin{equation}
p_k = (\beta-1) \frac{\Gamma(2\beta-3) \Gamma(k+\beta-3)}
{\Gamma(\beta-2) \Gamma(k+2\beta-3)} \propto \mbox{\rm} k^{-\beta}
\; (k \gg 1)
\label{krap}
\end{equation}
Similar results were obtained with other 
choices of $p_k$. 
\par
In Table 1 is given the average number of triangles at $G=0$, viz. 
$\langle T\rangle_{G=0}$, measured directly 
in Monte Carlo simulation (the 
left column) and estimated using eq. 
(\ref{T0}) (the right column) from the 
degree distribution generated in the same 
simulation for $\beta=2.5, 3$ and 
$4$, respectively. The agreement improves 
as $N$ increases and worsens, as 
expected, with decreasing $\beta$ and 
growing fat tail. Notice, that $\langle 
T\rangle_{G=0}$ increases with $N$ also 
for $\beta=4$, although it is expected 
to be finite in the limit $N \to \infty$: 
the second moment of the degree 
distribution is still rising 
significantly in the explored range of $N$.

\begin{figure}[ttt]
\vspace{1cm}
\includegraphics[width=6cm]{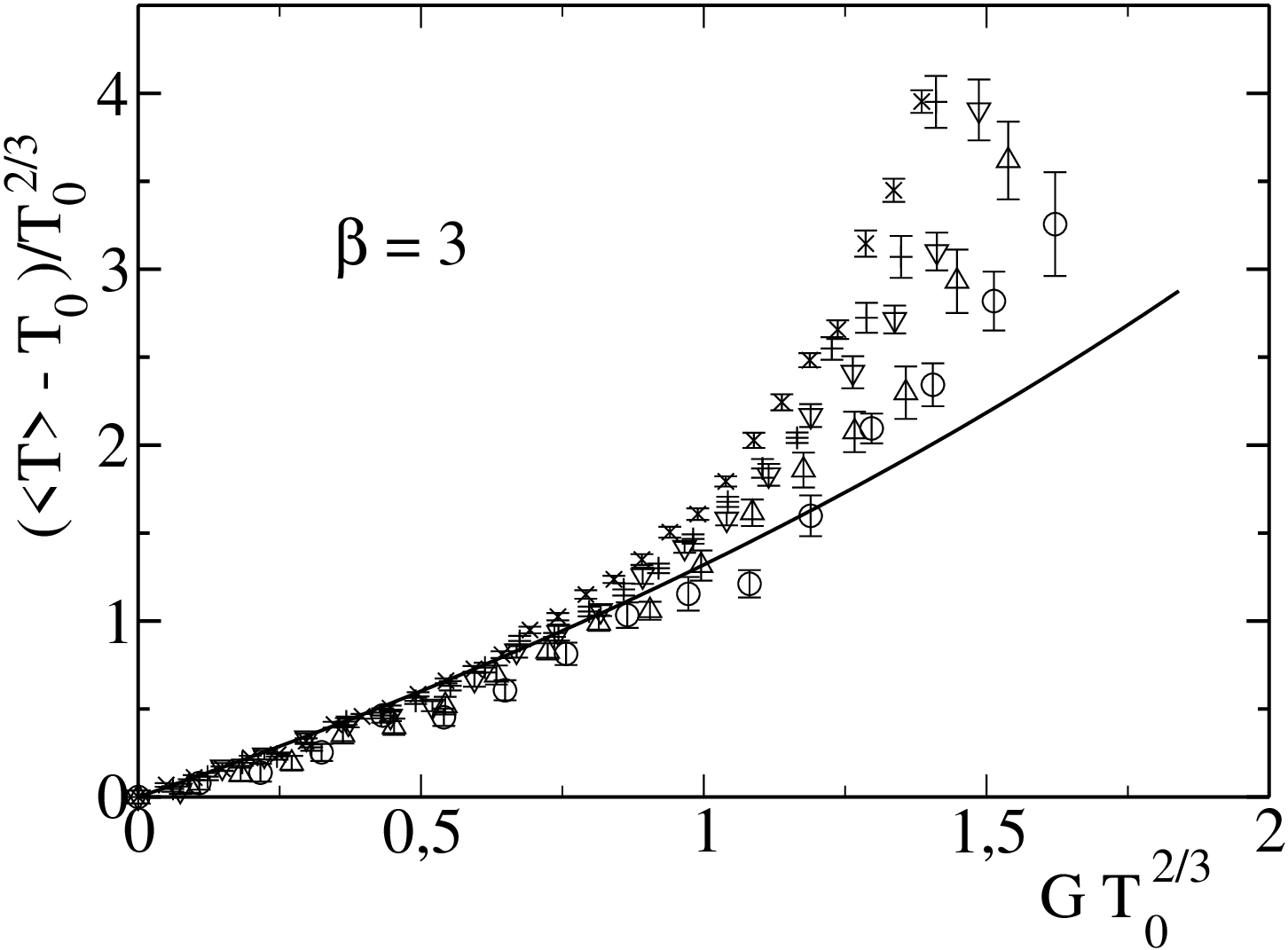}
\caption[Fig.4]{($\langle T\rangle-T_0)/T_0^{\frac{2}{3}}$ versus
$G T_0^{\frac{2}{3}}$ for $\beta=3$ and $N=2^{10}\; (\times)$,
$2^{11}\; (+)$, $2^{12}\; (\bigtriangledown)$,
 $2^{13}\; (\bigtriangleup)$, $2^{14}\; (\bigcirc)$.}
\end{figure}
\par
In Fig. 3 we plot $\langle T\rangle/T_0$ 
versus $G$ for $\beta=4$ and $N$ 
ranging from $2^{10}$ to $2^{14}$. The 
asymptotic expectation $\exp(G)$
is also drawn. The data scale and 
follow the curve $\exp(G)$ at small $G$. 
The characteristic fan shows up as 
one moves toward the transition point 
$G=G_{out}$. Notice, that largest $N$ 
data tend to be closer to the curve, 
as expected. At this point we are unable 
to tell how $G_{out}$ behaves. We 
recall that $G_{out} \propto \ln{N}$ 
when in zeroth order the graphs are of 
Erd\"os-R\'enyi type. This is expected 
to be the generic behaviour when the 
degree distribution has effectively 
a finite support. In this respect the 
status of the $\beta=4$ case is 
uncertain. There is no evidence in the data 
for an increase of $G_{out}$ with $N$. 
The apparent constancy of $G_{out}$ 
could be a finite-size effect, however. 
On the other hand, it is very plausible 
that hubs, \ie nodes with largest degree,  
behave as seeds of Strauss 
cliques, preventing the growth of the 
barrier separating the smooth phase 
from the crumpled one. In order to settle 
this question one would have to simulate 
enormous networks, beyond reach with 
present means.
\par
In Fig. 4 we plot 
$(\langle T\rangle- T_0)/T_0^{\frac{2}{3}}$
versus $GT_0^{\frac{2}{3}}$, as suggested by eq. (\ref{ttt}), 
for $\beta=3$ and $N$ ranging again from $2^{10}$ 
to $2^{14}$. The data scale 
reasonably well, especially at 
low $GT_0^{\frac{2}{3}}$. It appears that, 
very roughly
\begin{equation}
G_{out} \approx \frac{1.5}{T_0^{\frac{2}{3}}} \propto \ln^{-2}N
\end{equation}
The constant $c$ can be estimated from 
eq. (\ref{c}), using the observed (\ie cut)
degree distributions. A reliable estimate 
is obtained from low order moments 
and using rather large $N$ data; one 
finds $c$ ranging from $0.9$ to $1.2$. 
Setting $c=0.9$ one gets from eq. 
(\ref{ttt}) the curve shown in Fig. 4.
\par
In Fig. 5 we plot $\langle T\rangle/T_0$ 
versus $G T_0^{\frac{2}{3}}$, as 
suggested by eq. (\ref{uuu}), for $\beta=2.5$ 
and $N$ ranging from $2^{10}$ to 
$2^{14}$. One observes the expected 
scaling at small values of $G T_0^{\frac{2}{3}}$, 
but at larger values finite-size effects become gradually
more and more important. It appears that, very roughly
\begin{equation}
G_{out} \approx \frac{2.3}{T_0^{\frac{2}{3}}}
 \propto N^{-\frac{1}{2}}
\end{equation}

\begin{figure}
\includegraphics[width=6cm]{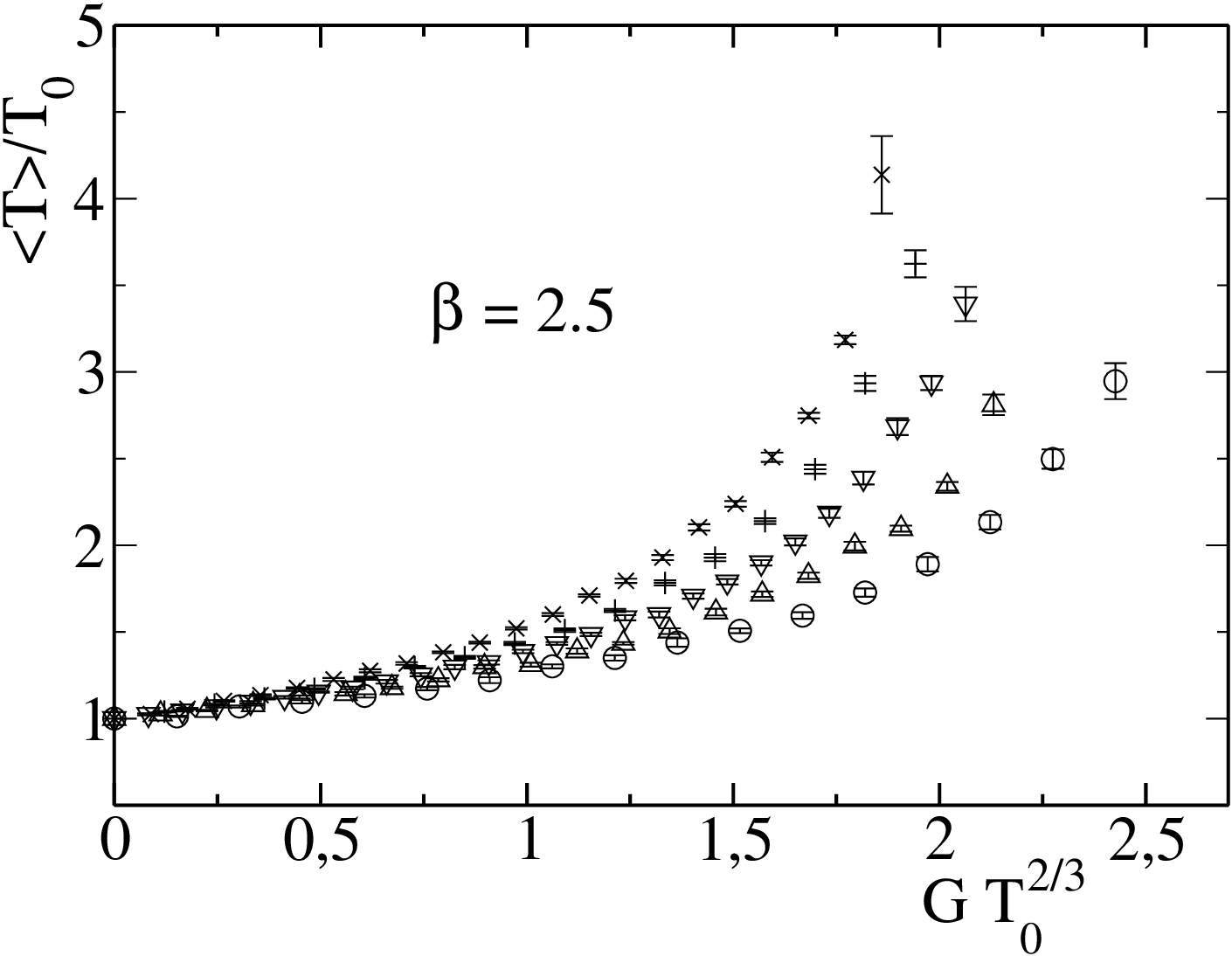}
\caption[Fig.5]{$\langle T\rangle/T_0$ versus
$G T_0^{\frac{2}{3}}$ for 
$\beta=2.5$ and $N=2^{10}\; (\times)$,
$2^{11}\; (+)$, $2^{12}\; (\bigtriangledown)$, 
$2^{13}\; (\bigtriangleup)$,
$2^{14}\; (\bigcirc)$.}
\end{figure}
\par
In Fig. 6 we plot the degree distribution 
at $N=2^{14}$ and $\beta=2.5$, 
for $G=0$ and $0.08$ ( just before 
the transition to the crumpled phase), 
to show that the introduction of the 
interaction Hamiltonian does not spoil 
the scale free property. In Fig. 7 we 
show the variation with $k$ of the 
clustering parameter 
\begin{equation}
C_k  = \frac{2 T_k}{k(k-1)}
\end{equation}
where $T_k$ is the average number of 
triangles touching a vertex with degree $k$.
One observes that the decrease of the 
clustering parameter is much slower than 
observed in Internet, for example, 
where $C_k \propto k^{-0.75}$ \cite{vpv}.

\begin{figure}
\vspace{0.6cm}
\includegraphics[width=6cm]{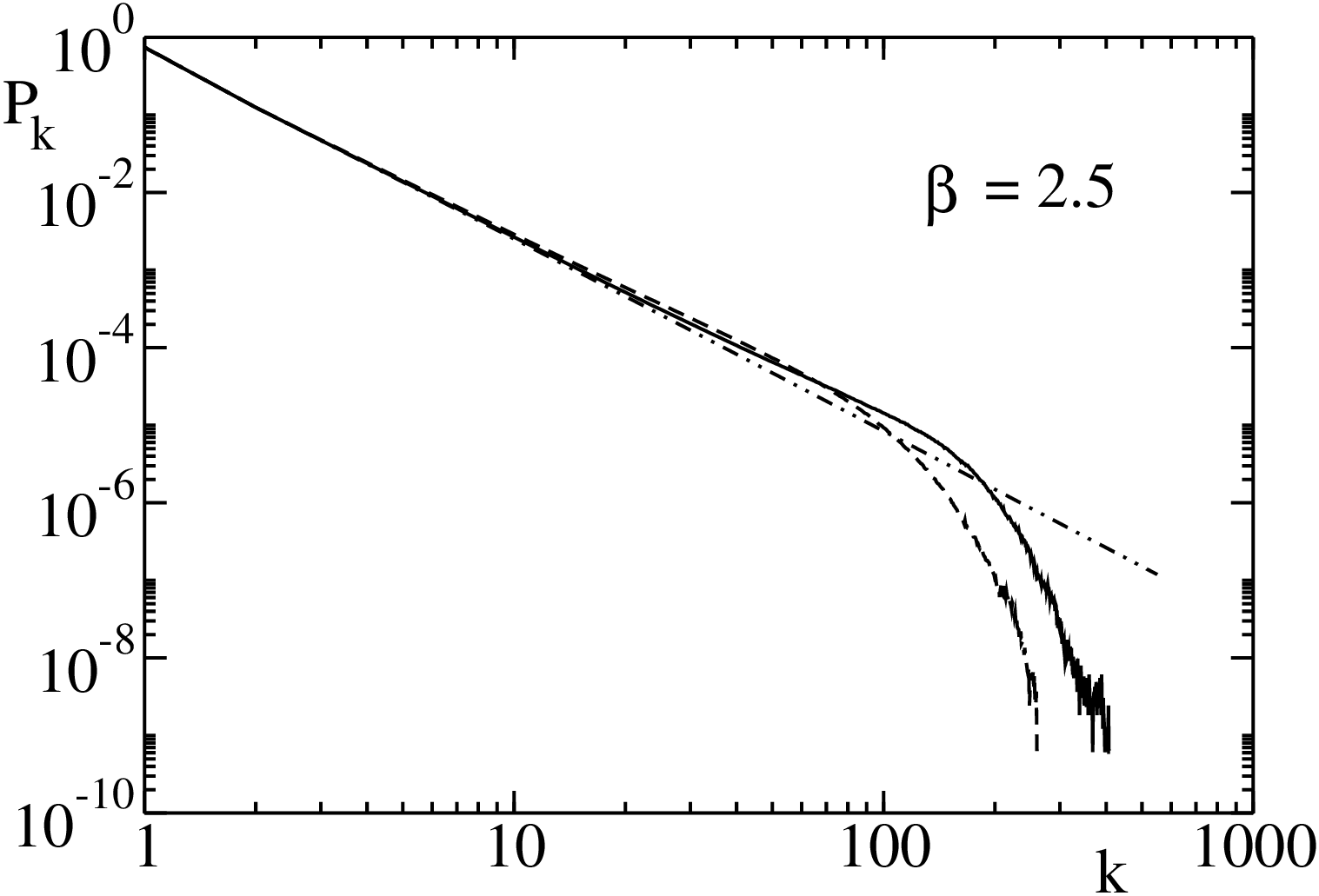}
\caption[Fig.6]{The degree distribution 
$P_k$ at $N=2^{14}$ and for
$\beta=2.5$. Dashed line is for $G=0$, 
solid line for $G=0.08$
(just before the transition to the 
crumpled phase). The almost straight
line corresponds to the asymptotic 
shape of $P_k$, \ie to $p_k$.}
\end{figure}

\begin{figure}
\vspace{1cm}
\includegraphics[width=6cm]{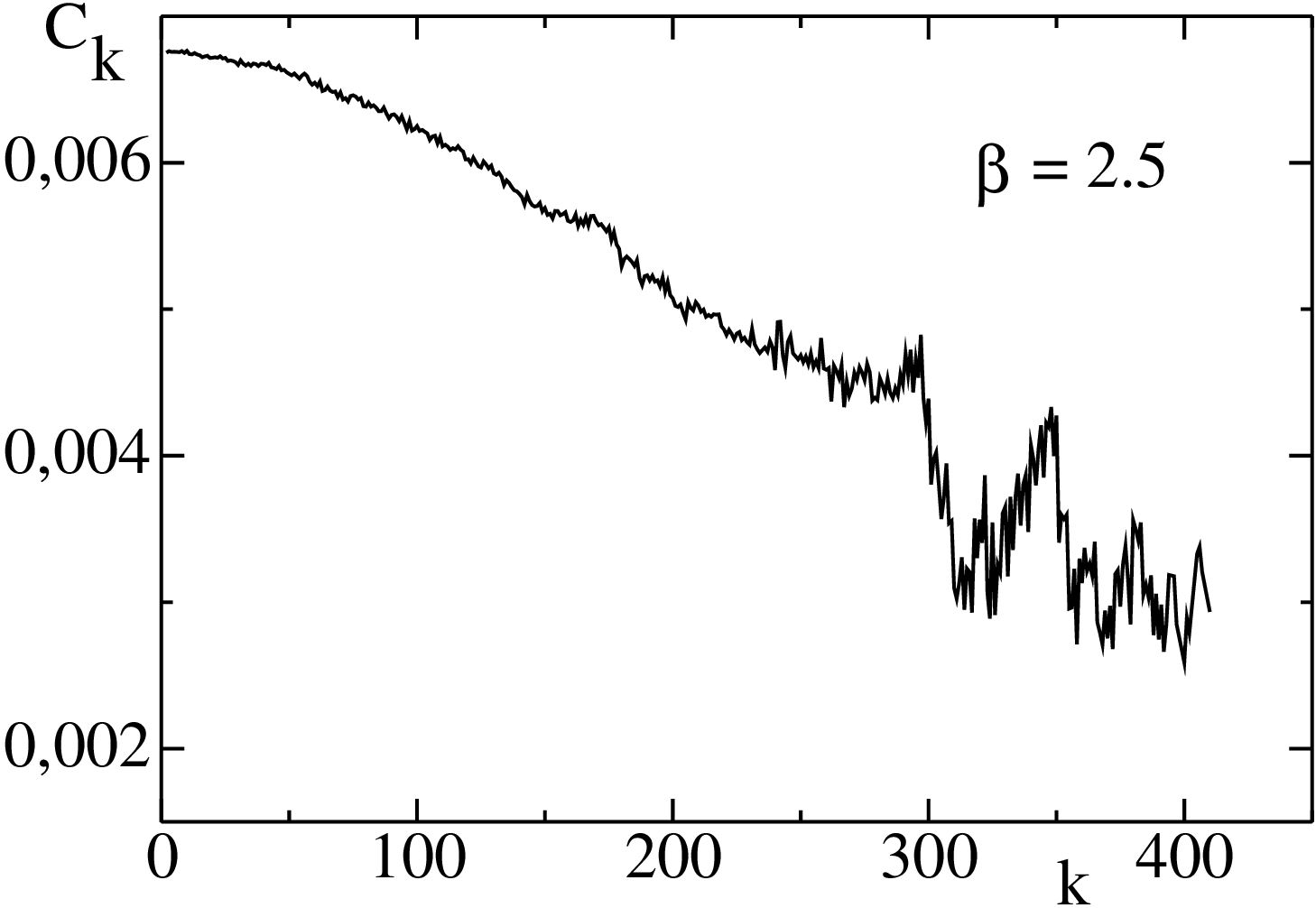}
\caption[Fig.7]{Clustering coefficient 
versus node degree at $N=2^{14},
\beta=2.5$ and $G=0.08$ (just before 
the transition to the crumpled phase). 
Notice, that the scales are linear.}
\end{figure}

\section{SUMMARY AND CONCLUSION}
This paper is a direct continuation of 
ref. \cite{bjk}, where we have studied 
perturbed random Erd\"os-R\'enyi graphs.   
In the present paper we show that 
the results of ref.\cite{bjk} continue to hold when 
an ensemble of (almost) arbitrary 
uncorrelated graphs is perturbed by the 
same interaction, favouring the formation 
of triangles. There is, however, a
notable exception to this generic 
behaviour: the so-called scale-free graphs
behave differently, especially when the 
degree distribution has a diverging 
variance.
\par
At finite $N$ the smooth phase exists 
only when the interaction coupling $G$ 
is smaller than some threshold $G < G_{out}$. 
Generically $G_{out}$ scales like
$\ln{N}$, but for scale-free networks 
with $2 < \beta < 3$ it scales like
$N^{\beta-3}$ (modulo logs). Hence, 
the support of the smooth regime does 
not expand but shrinks to zero in 
the thermodynamic limit. There is nothing 
dramatic in such a behaviour, which is
also encountered in more conventional matrix 
models. It just means that $G$ is not
the physical coupling and that the latter 
is rather $G N^{3-\beta}$, again modulo logs.  
When the physical coupling is used the smooth 
phase lives in a finite coupling
interval and the average number of triangles 
is enhanced compared to the unperturbed expectation. 
However the stability of this phase 
is an open problem. In ref.
\cite{bjk} we found that nonperturbative 
phenomena are manifestly negligible
almost in the whole smooth phase: numerical data 
were remarkably close to
perturbative predictions. Here, we cannot 
claim the same, partly because we are
unable to disentangle finite-size and 
nonperturbative effects.
\par
At finite $G$ scale-free networks are
unstable for $N$ large enough. The physical 
significance of this singular behaviour  
is not fully clear yet. In our 
simulations the thermal motion consists of
network rewirings. Rewiring is an 
ergodic move: every two states can be
transformed one into another by making 
a finite number of rewirings. In
particular, any other algorithmic move 
could be regarded as made up of
rewirings. However, with a different 
algorithm the thermalization 
time of the system would in general 
be different, in particular it could 
explode with increasing $N$. Thus, it 
is not excluded, although does not 
seem very likely, that the instability 
is an algorithm artifact.
\par
Very many natural networks are scale-free, 
with the exponent $\beta$ below 3.
At least some of them seem fairly stable. 
In some cases there are selection rules 
constraining the rewirings. But this 
is not the most interesting possibility. 
How do the natural networks compare 
to the graphs of our model? We see one 
significant difference: in our graphs 
the clustering coefficient is weakly 
correlated with node degrees, while 
in natural networks it tends to decrease 
like some power of the latter. The behaviour 
of our graphs is easy to understand 
as follows: The system forms triangles 
at random and therefore tends to 
attach many triangles to hubs, 
which apparently become seeds of 
Strauss cliques. Natural networks 
seem to avoid this disease by suppressing 
the formation of triangles at hubs. 
It is plausible that a specific 
hierarchical organization screens natural 
networks from the instability. In such a
scenario a triangle generating term with a
finite coupling $G$ could after all be present 
in the Hamiltonian. For the moment
this is just a speculation.
\par
Our paper is not a phenomenological one.
We are not yet at the stage of constructing
a model to be compared to the data. We
focus on the theoretical problem of the
stability of networks with respect to motif
generating terms in the Hamiltonian.
However, the paper is not quite devoid of
phenomenological implications: our method allows us
not only to calculate averages of physical quantities
characterizing an individual network, but also fluctuations
of those quantities in the ensemble, giving us an insight
into the problem of typicality of networks. As far as we know 
the magnitude of fluctuations of motifs has never
been estimated for graphs with an arbitrary given degree
distribution.

\vspace{0.5cm}
\begin{center}
{\bf ACKNOWLEDGMENTS}
\end{center}
This work was partially supported by the EC IHP Grant
No. HPRN-CT-1999-000161, by the Polish State Committee for
Scientific Research (KBN) grants 2P03B-08225 (2003-2006) 
and 2P03B-99622 (2002-2004), and by EU IST Center of Excellence 
"COPIRA". In the numerical calculations we used a PC farm located in 
the M. Smoluchowski Institute of Physics and donated by the 
Polish Science Foundation.  Laboratoire de Physique 
Th\'eorique is Unit\'e Mixte du CNRS UMR 8627.


\begin{thebibliography}{99}
\bibitem{foot} The rules to find the expectation value of the string 
of a given diagram can be derived elegantly using the hidden 
variable formulation of network models. Without entering 
into details, we outline the idea: Let the letter $h$ denote the random 
hidden variable associated with a node and let the probability of 
connecting two nodes be $f(h)f(h')/N$. Using the formalism of the 
model one finds
$$ \langle k!/(k-m)!\rangle = \langle f(h)\rangle^m
\langle f(h)^m\rangle $$
and, in particular, $\langle k\rangle =\langle f(h)\rangle^2$. Using 
this result and the fact that with each vertex, say $a$, of the diagram 
is associated a factor $\langle f(h)^{m_a}\rangle/N^{m_a/2}$, one 
easily recovers the rules given in the main body of the text.
\end{thebibliography}
\end{document}